\documentclass[twocolumn]{webofc}

\usepackage[varg]{txfonts}   
\usepackage{hyperref}
\usepackage{url}
\hypersetup{colorlinks=true,citecolor=blue,urlcolor=blue,linkcolor=blue}
\begin{document}
\title{Model Independence of Effective Field Theory}

\author{\firstname{Anthony} \lastname{Thomas}\inst{1}\fnsep\thanks{
\email{anthony.thomas@adelaide.edu.au}} }

\institute{CSSM, Department of Physics, The University of Adelaide, Adelaide SA 5005, Australia }

\abstract{The use of effective field theory offers a systematic way to improve calculations of nuclear reactions and the properties of atomic nuclei. Its successes have led to the widespread belief that the predictions of this approach are model independent. We explain why this is definitely not the case.
}
\maketitle
\section{Introduction}
\label{intro}
In a classic paper written in 1990, Weinberg proposed a new approach to nuclear forces~\cite{Weinberg:1990rz} based upon the approximate chiral symmetry of QCD~\cite{Gasser:1983yg,Thomas:2025wlj}. Many authors have developed this approach with considerable success -- see for example Refs.~\cite{Ordonez:1995rz,Machleidt:2024bwl,Epelbaum:2019kcf,Phillips:2021yet}. 

The idea is to calculate one- and two-pion exchange contributions to the nucleon-nucleon force using chiral perturbation theory to an order as high as possible. Discrepancies with the experimental phase shifts are then removed by adding contact terms to the Lagrangian density. With typically 20-30 parameters one achieves a description of the scattering data comparable in accuracy to that obtained using earlier phenomenological potential models. When the resulting potential is applied to the binding energies of light nuclei, the discrepancies are used to determine parameters for a three-body force; once again with its long-range part calculated using chiral perturbation theory. Additional details of this three-body force can be pinned down using other observables, such as neutron-deuteron scattering.

With its roots in a vital symmetry of QCD and its systematic approach to calculations, there appears to be a widespread belief that the approach is model independent. For example, it is common to talk about determining {\em the} three-body force in this way. 

\begin{figure*}
\centering
\vspace*{1cm}       
\includegraphics[width=15cm,clip]{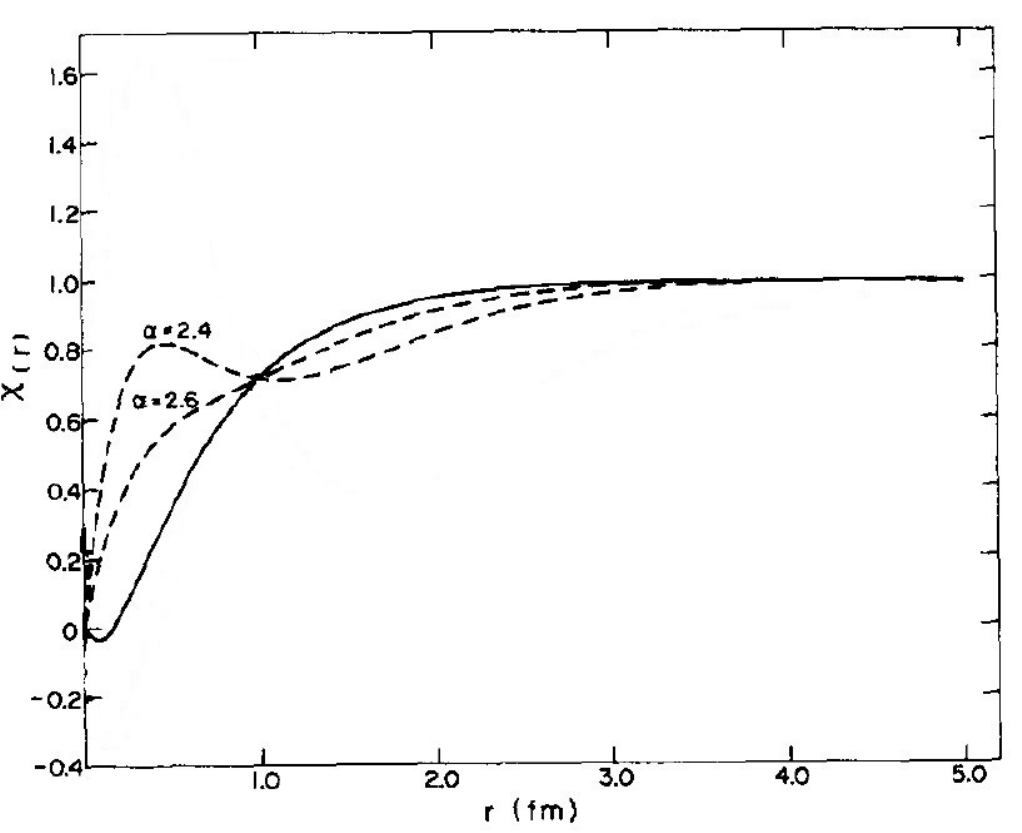}
\caption{Changes in the short distance wave function in the $^1S_0$ channel at zero energy before the unitary transformation (solid) and afterwards (dashed), with two choices of the parameter $\alpha$. The range of $\alpha$ is chosen so as to not change the one-pion-exchange tail of the nuclear force.}
\label{fig1}       
\end{figure*}
My concern about this point of view is not simply a matter of semantics. Rather, it goes to the heart of what we really know about the strong force. For example, when the systems of interest involve higher densities, for example in the core of neutron stars, the role of the three-body force is crucial. It would be reassuring if, indeed, the three-body force found within the chiral effective field theory were unique. One could then have considerable faith in the reliability of the predicted equation of state. 

Unfortunately, this is not the case. There is an implicit assumption in this approach that the short-distance nucleon-nucleon force is local. Given that the nucleon itself is a complicated structure built of quarks and gluons, with a size comparable to the inter-nucleon separation in a nucleus, this assumption seems unlikely.

Our purpose in this short contribution is to recall work carried out some time ago, which demonstrates clearly that fitting nucleon-nucleon phase shifts over a wide range of energies is not enough to ensure that the predictions for nuclear properties obtained using that potential are model independent. Indeed, by making a unitary transformation on the corresponding Hamiltonian, one can generate an infinite number of two-body interactions, which fit exactly the same phase shifts, while yielding different results for nuclear binding energies. Clearly, if the two-body predictions vary then so will the three-body forces needed to restore agreement with nuclear data. Then the three-body force itself is model dependent and so are the predictions, for example, for dense matter.

In the next section we recall the details of one particular unitary transformation and the effects on the predictions for the binding energy of the triton and the neutron-deuteron doublet scattering length. The final section includes some closing remarks.

\section{A unitary transformation}
\label{sec-1}
In this section we briefly outline the calculations of Afnan and Thomas~\cite{Thomas:1975qfg}. We start with the Hamiltonian $H \, = \, K \, + \, V_N$, which for simplicity we choose to describe nucleon-nucleon scattering in the $^1S_0$ channel. We then make the unitary transformation
\[
\tilde{H} = U H U^{\dagger} = K + \tilde{V}_N \, ,
\] 
with the specific choice
\[
U = 1 - 2 | h \rangle \langle h | \, ,
\]
where 
\[
\langle h | h \rangle = 1
\]
and 
\[
\langle r | h \rangle = N e^{-\alpha r} (1 - \beta r)
\, . \]
\begin{figure*}
\centering
\vspace*{1cm}       
\includegraphics[width=15cm,clip]{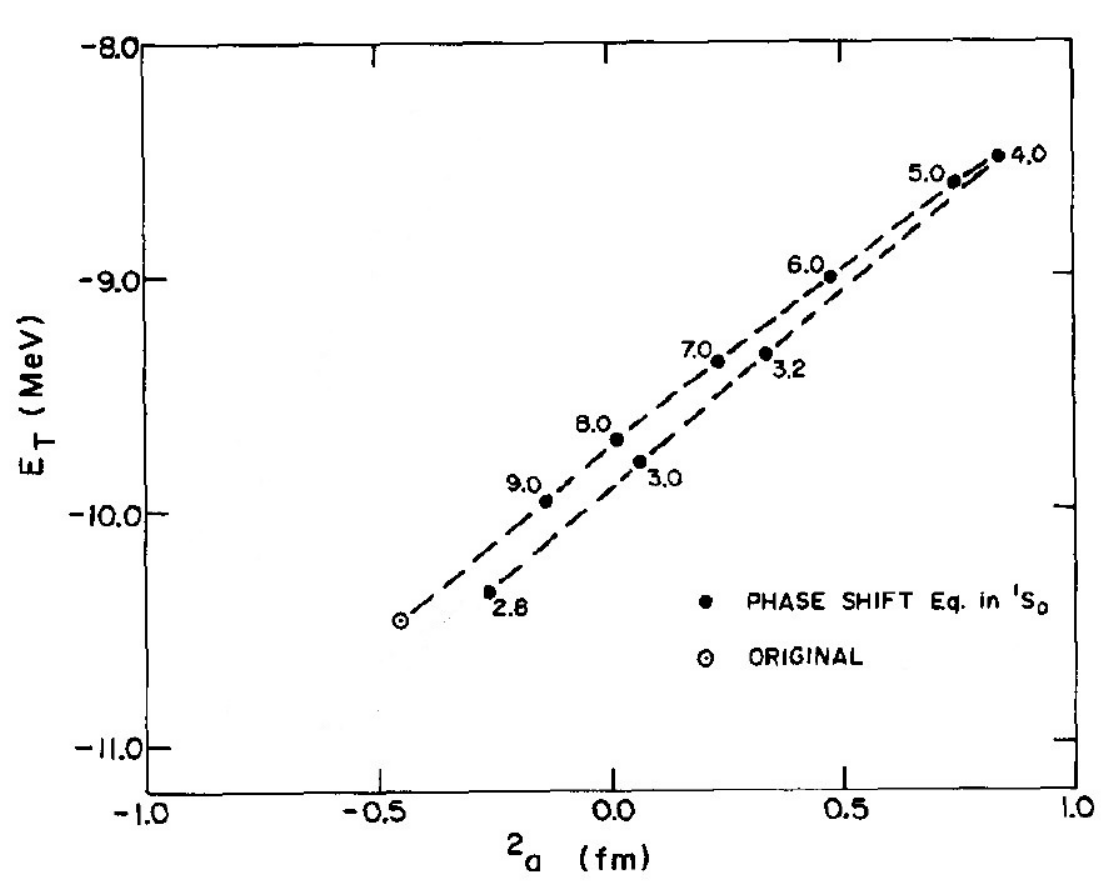}
\caption{The results of Afnan and Thomas~\cite{Thomas:1975qfg} for the variation of the triton binding energy and the spin-1/2 neutron-deuteron scattering length as a function of the parameter $\alpha$ entering the unitary transformation described in the text.}
\label{fig2}       
\end{figure*}
The parameter $\beta$ was chosen to be 1 fm$^{-1}$ and $\alpha$ needed to be sufficiently large that the long-range behavior of the scattering wave function and hence the phase shifts were not changed. Writing the scattering wave function in the $^1S_0$ channel as  
\[ \psi_k(r) \, = \, j_0(kr) \, + \, \frac{\chi(r)}{r} \, ,
\]
we show the component $\chi(r)$ in Fig.~\ref{fig1}, at zero energy, for two values of $\alpha$. By restricting $\alpha$ to values larger than 2.6 fm$^{-1}$, we ensure that the behavior of the wave function associated with the one-pion exchange tail of the nucleon-nucleon interaction would not be altered. All changes are restricted to short distances. This choice also guarantees that charge symmetry is respected in that the Coulomb-modified proton-proton scattering length, $a_{pp}$, remains with 5\% of that for the untransformed potential.

The results we show in the following were obtained using the two-term Mongan (case II) separable potential~\cite{Mongan:1969dc} in the $^1S_0$ and $^3S_1 - ^3D_1$ partial waves. Of course, one could include higher partial waves but their total contribution to the triton binding energy is minor. We could also have applied the unitary transformation to the triplet partial wave but the point we wish to make is illustrated perfectly adequately by transforming just the singlet case.

Two key observables for the low-energy neutron-deuteron system are the binding energy of the triton and the doublet (spin-1/2) scattering length. For some time it has been known that the results for all two-body potentials which fit nucleon-nucleon scattering data lie on a straight line, called the Phillips 
line~\cite{Phillips:1968zze}. 
This also holds true for the new non-local potentials generated by our unitary transformations, as long as $\alpha$ is larger than 2.4 fm$^{-1}$. This is illustrated in Fig.~\ref{fig2}. 

From Fig.~\ref{fig2} we see that the {\em variation in the binding energy of the triton is as much as 2 MeV, or 20\% of its experimental value, for a selection of two-body potentials which produce exactly the same phase shifts at all energies}. It is then clear that if one were to choose the contact pieces of the three-body force such that it reproduces the experimental binding energy of the triton, their variation, and hence the model dependence of the three-body force, would be very large indeed.

\section{Concluding remarks}
By considering exact solutions of the three-body problem, we have provided a simple illustration of the model dependence of the three-body force. Once one realizes that there is no reason for the short-distance nucleon-nucleon potential to be local, even fitting the phase shifts over all energies from zero to infinity is not sufficient to guarantee potentials which yield the same results in a three-body system. Similar results have been reported for infinite nuclear matter~\cite{Coester:1970ai}. Since the three-body force is specified by fitting the difference between experiment and the predictions using a two-body potential, the model dependence follows.

At the present time, there is considerable interest in the equation of state (EoS) of dense matter; in the case of neutron stars this requires that $\beta$-equilibrium be imposed, while this is not the case for high-energy heavy-ion collisions. The argument that we have presented should serve as a reminder that one should keep an open mind when it comes to models of the EoS of dense nuclear matter. In particular, differences between predictions of models such as the quark meson coupling (QMC) model~\cite{Martinez:2020ctv,Guichon:2018uew,Stone:2019blq}, with which I have been involved, should be evaluated against their capacity to describe experimental data, rather than their agreement with chiral nuclear forces.

\section*{Acknowledgement}
It is a pleasure to thank Iraj Afnan for his comments on this manuscript and to gratefully acknowledge his collaboration on this and numerous other projects over many years. I would also like to thank Ross Young for a helpful discussion concerning the issues raised in this manuscript. This work was supported (through CSSM) by the University of Adelaide, as well as by the Australian Research Council through Discovery Project DP230101791.

\bibliography{Refs}


\end{document}